\documentclass[doublecol]{epl2}

\usepackage{amsmath,amssymb}
\usepackage{longtable}
\usepackage{graphicx}
\def\aeti{$\alpha$-(BE\-DT\--TTF)$_2$\-I$_{3}$}
\def\cm{cm$^{-1}$}

\title{Effect of electronic correlations on the metal-insulator transition of $\alpha$-(BEDT-TTF)$_2$I$_3$:
theoretical and experimental investigations of its optical properties}
\shorttitle{Effect of electronic correlations on the metal-insulator transition of $\alpha$-(BEDT-TTF)$_2$I$_3$}


\author{T. Peterseim \and M. Dressel}

\institute{
  1.~Physikalisches Institut, Universit\"at Stuttgart, Pfaffenwaldring 57, 70550 Stuttgart, Germany}

\date{\today}

\abstract{
The organic salt $\alpha$-(BEDT-TTF)$_2$I$_3$ is considered a model system for metal-insulator transition
due to electronic charge ordering at $T_{\rm CO}=135$~K. The optical properties obtained from polarized reflection measurements above and below $T_{\rm CO}$ can be well described by calculations based on first-principle density-functional theory (DFT). We discuss the effect of electronic correlations on the metal-insulator transition.}

\begin{document}
\maketitle

\section{Introduction}
The charge-transfer salt \aeti\ was the first molecular crystal to show
high-conductive electronic properties in two dimensions\cite{Bender84a,Bender84b,Dressel94}.
Fukuyama and collaborators\cite{Kino95,Kino96,Seo04} proposed that
the pronounced metal-to-insulator transition at $T=135$~K
is caused by the development of a charge-ordered state driven by electronic correlations.
Subsequently, the emergence of a second doublet in $^{13}$C-NMR spectra\cite{Takano01,Takahashi06},
the splitting of charge-sensitive modes in Raman-\cite{Wojciechowski03,Yakushi12} and
IR-spectroscopy\cite{Ivek11,Beyer16}, and careful
x-ray diffraction measurements\cite{Kakiuchi07} unambiguously confirmed that the
system establishes charge disproportionation.

\begin{figure}
    \centering
       \includegraphics[width=0.99\columnwidth]{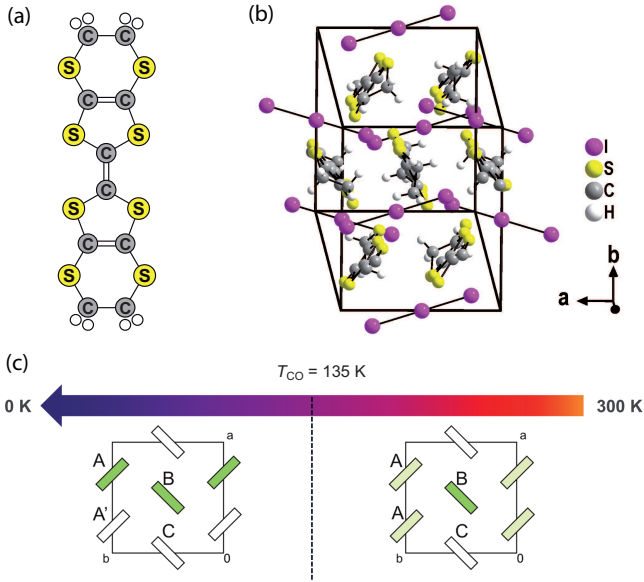}
    \caption{(a)~Sketch of the bis-(ethyl\-ene\-di\-thio)\-te\-tra\-thia\-ful\-va\-lene
molecule, called BEDT-TTF. (b)~Drawing of the \aeti\ unit cell with orientation along the $c^{*}$-direction. The unit cell contains four BEDT-TTF and two I$_3$ molecules. The cations are arranged in a herringbone-like structure within the $ab$-plane separated from each other along the $c^{*}$-direction by the anions. (c)~Illustration of the development of the charge disproportionation and crystal symmetry in \aeti\ when cooling from room temperature to 0~K. The $ab$-plane is displayed with the four BEDT-TTF molecules labeled by A, B, and C;
the anions are not included for simplicity. The A molecules are related to each other by a point inversion on their connecting line. The opacity of the green color represents the charge imbalance between the molecules. At $T=300$~K there is an imbalance between molecule B and C. Below $T_{\rm CO}$ the point inversion is lost and the A molecules become inequivalent. A stripe-like charge-order pattern develops along the $b$-direction.
    \label{fig:structure}}
\end{figure}

The title compound consists of layers formed by the BEDT-TTF molecules [fig.~\ref{fig:structure}(a)] that are separated along the $c^{*}$-direction by I$_3^-$ anions creating a two-dimensional system.
Each anion receives one electron from two BEDT-TTF molecules leading to half an electronic charge per cation. Thus, the system can be considered as a quarter-filled system. The room temperature conductivity in the $ab$-plane is rather high, ranging between $10$ and $100~(\Omega {\rm cm})^{-1}$, depending on the specimen. The resistivity ratio $\rho_{a}:\rho_{b}=2:1$ indicates the quasi-two-dimensional character of this material.
In contrast, the resistivity perpendicular to the $ab$-plane is  larger by a factor 1000. The intuitive picture of holes as dominant charge carriers in the BEDT-TTF layers is confirmed by thermopower and Hall effect measurements\cite{Bender84a,Pokhodnya87,Tajima00}.
Previous theoretical calculations\cite{Emge86,Ishibashi06,Mori10,Alemany12},   however, consistently yield
a semi-metallic behavior at ambient conditions: the Fermi surface intersects the valence band  and conduction band leading to hole and electron pockets.

\begin{figure}
    \centering
    \includegraphics[width=0.9\columnwidth]{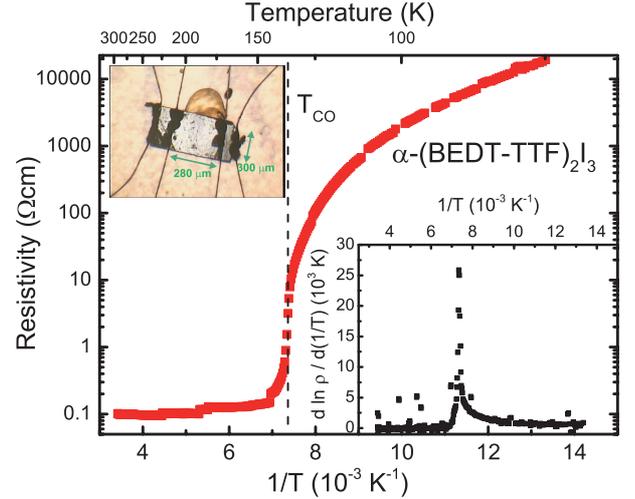}
    \caption{Resistivity of \aeti\ as a function of inverse temperature $1/T$ between 300 and 75~K. The dashed vertical line marks the metal-insulator transition at $T_{\rm CO}=135$~K, which is characterized by an abrupt and steep increase of $\rho(T)$ by several orders of magnitude. In the inset the derivative of the logarithm of the resistivity $\rho(T)$  is plotted as a function of $1/T$. The transition manifests itself as a sharp peak indicating a first-order transition. The photograph shows the cut \aeti\ crystal with gold wires attached by carbon paste. The sample was glued to the ground plate by GE vanish. The distance of the inner contacts is $280~\mu$m  and the width was about $300~\mu$m, the sample thickness is approximately $80~\mu$m. \label{fig:Resistivity}}
\end{figure}

On cooling the resistivity shoots up by several order of magnitudes in a narrow temperature range around 135~K as depicted in fig.~\ref{fig:Resistivity}. Heat-capacity measurements also evidence this first-order metal-insulator transition with a diverging behavior at the transition temperature\cite{Fortune91}.  NMR and optical experiments as well as x-ray scattering studies \cite{Takano01,Takahashi06,Wojciechowski03,Yakushi12,Ivek11,Beyer16,Kakiuchi07} render charge disproportionation between the molecular sites indicating the development of a long-range horizontal charge order below $T_{\rm CO}=135$~K.
However, signatures of weak charge disproportionation were also observed above the transition.

In general, charge order is due to electron-electron interactions in the form of
on-site Coulomb interaction $U$ in combination with nearest-neighbor interaction $V$ as the driving forces for the metal-indulator transition\cite{Seo06}. Recently, density function theory (DFT) calculations on the basis of general gradient approximation (GGA) functional could determine both the metallic and the insulating state as well as the charge imbalance between the molecular sites\cite{Alemany12}. Since these model do not take into full account electronic correlations, it became questionable how much electron-electron interactions actually influence the physical properties. In fact, they found the changing strength of the donor-anion hydrogen bonding interaction is crucial for the redistribution of holes on the BEDT-TTF molecules.

Besides dc resistivity, optical properties are most sensitive to the metal-insulator transition and charge disproportionation, as they reflect the local charge distribution in vibrational features\cite{Dressel04,Drichko09,Girlando11,Yakushi12} as well as the electronic response in a large dynamic range\cite{Dressel04}.
Although the infrared optical properties of \aeti\ have been measured
repeatedly\cite{Koch85,Sugano85,Meneghetti86,Yakushi87,Zelezny90,Dressel94,Clauss10,Yue10,Ivek10,Ivek11,Beyer16},
no attempt has been reported, to link the experimental results with the calculated band structure.
Hence, we have performed extended  calculations based on density functional theory in order to determine the band dispersion of \aeti\ along a specific path as well as the electronic excitation, which we subsequently compare directly  to our experimental results. We conclude that correlation effects are not as important to drive the system into the charge-ordered phase as previously presumed.

\section{Electronic band structure}
\label{sec:bandstructure}
The band structure of \aeti\ was determined by \textit{ab-initio} DFT calculations as standardly implemented in the software package Quantum Espresso (Version 4.3.2 and 5.1)\cite{Giannozzi09}. In contrast to the well-established H{\"u}ckel theory with its molecular orbitals, the semi-empirical DFT is based upon plane-waves and uses the advantage of the crystal periodicity. We employed a norm-conserving PBE general gradient approximation (GGA) functional \cite{Perdew96} for all atom types, up to a certain level taking into account the exchange correlation as well as the spatial variation of the charge density.
The cut-off energy for the plane waves and electronic density was set to 30~Ry and 120~Ry, respectively. The self-consistent energy calculations were performed on a regularly spaced grid of $8\times8\times4$ Monkhorst grid\cite{Monkhorst76}. Since \aeti\ is metallic at room temperature, a smearing factor of 0.05~Ry was selected. The crystal structures determined from x-ray scattering experiments at room temperature and low temperature were taken from Emge {\it et al.} \cite{Emge86} and Kakiuchi {\it et al.} \cite{Kakiuchi07}. They are used without any optimization of the unit cell parameters or the atomic positions.

\begin{figure}
    \centering
    \includegraphics[width=0.9\columnwidth]{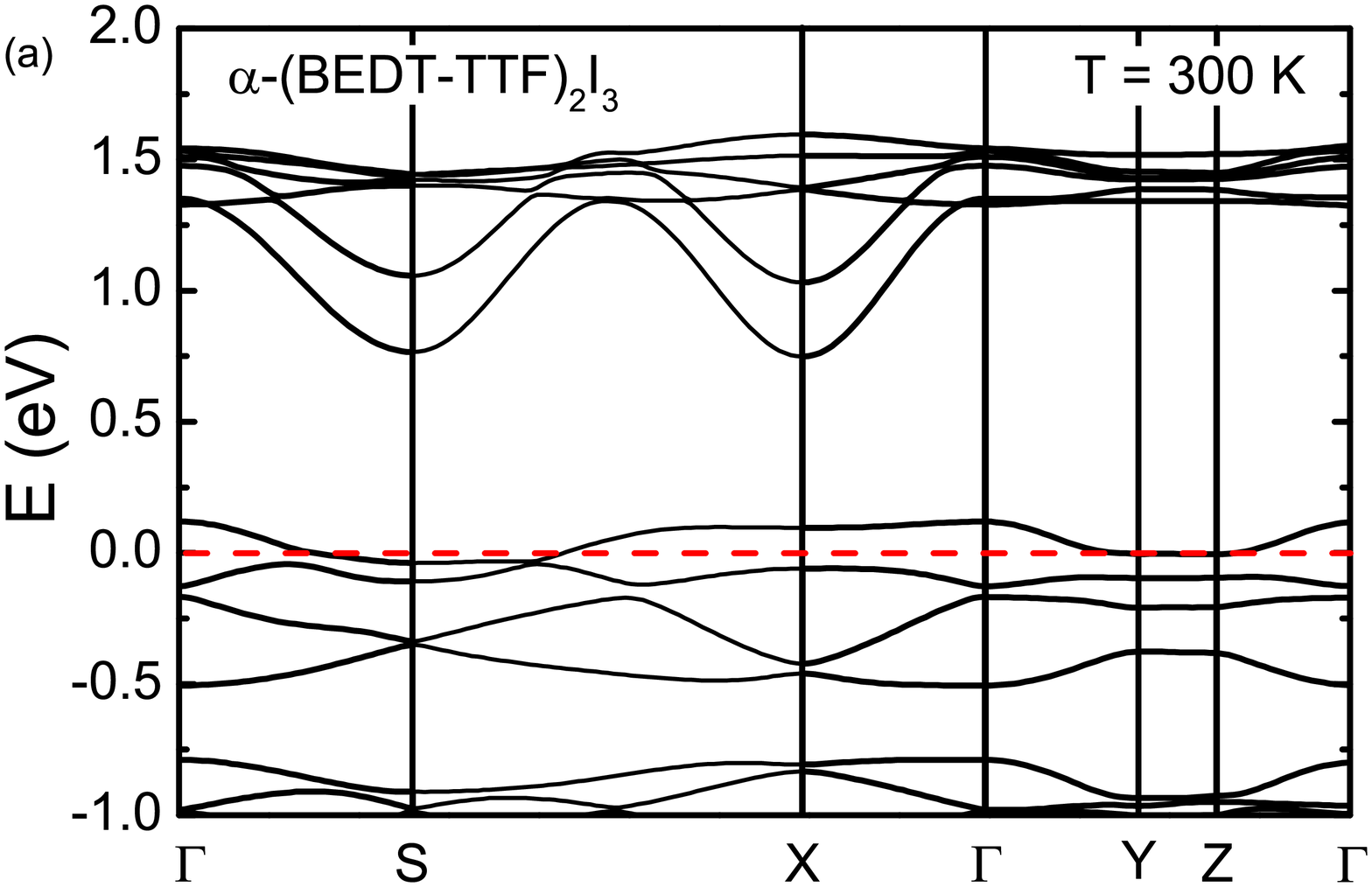}
    \includegraphics[width=0.9\columnwidth]{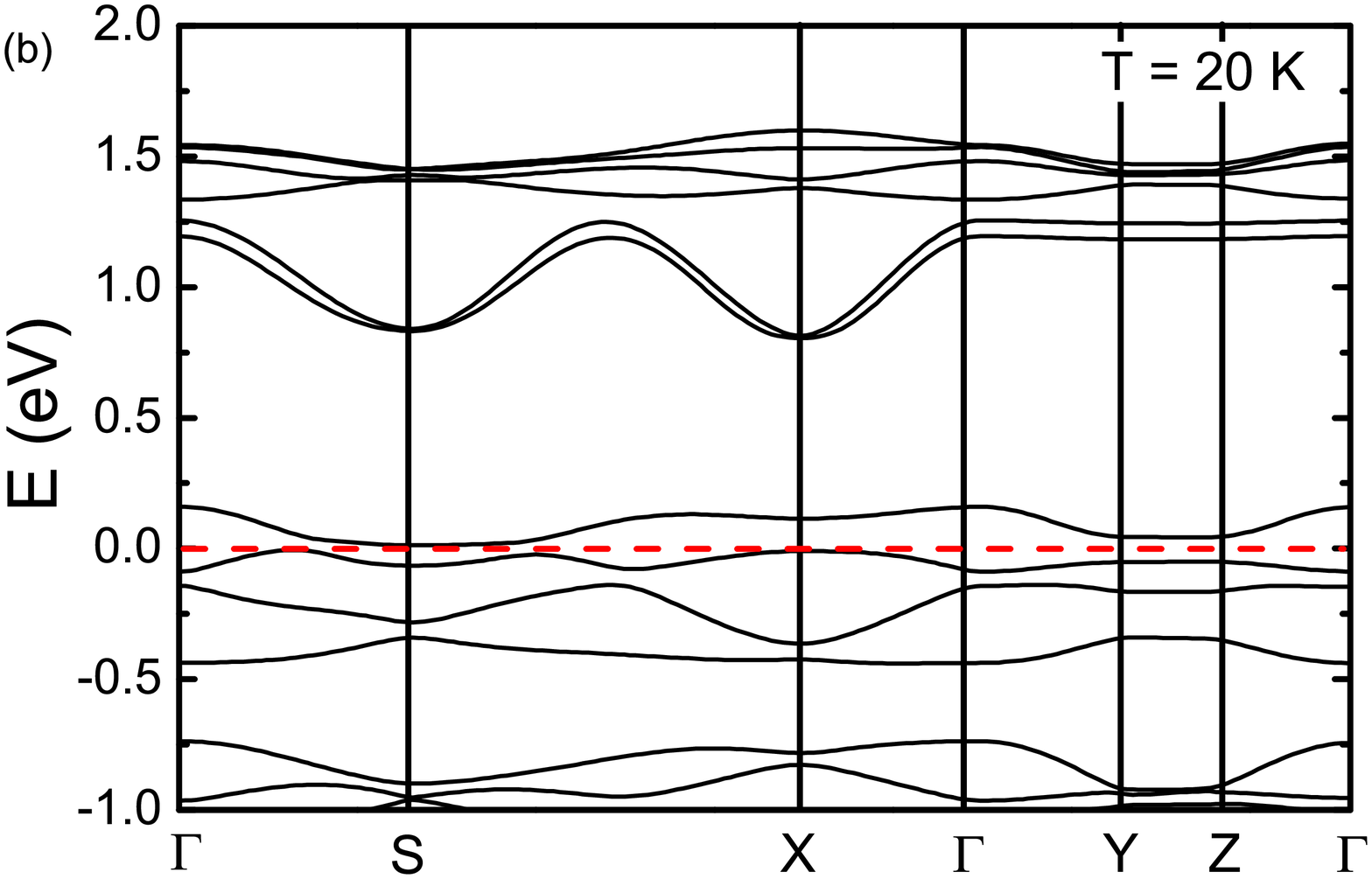}
    \caption{Calculated band structure of \aeti\ (a) in the metallic state at  $T=300$~K and (b) in the insulating state at $T=20$~K along the $k$-path: $\Gamma (0,0,0) \rightarrow  S (-0.5,0.5,0)\rightarrow  X (0.5,0,0) \rightarrow \Gamma (0,0,0) \rightarrow Y (0,0.5,0) \rightarrow \text{Z} (0,0,0.5) \rightarrow \Gamma (0,0,0) $. The Fermi level $E_F$ is indicated by the red dashed lines. \label{fig:Bandstructure}}
\end{figure}

The electronic band structure of \aeti\ is depicted in
fig.~\ref{fig:Bandstructure} covering
an energy range from -1 to 2~eV along the path $\Gamma (0,0,0)\rightarrow S (-0.5,0.5,0)\rightarrow  X (0.5,0,0) \rightarrow \Gamma (0,0,0) \rightarrow Y (0,0.5,0) \rightarrow \text{Z} (0,0,0.5) \rightarrow \Gamma (0,0,0)$ in units of the triclinic reciprocal lattice vectors. Three regions with several bands can be identified in the displayed energy range. The upper band located between 1~eV and 1.5~eV are attributed to the LUMO bands of the {BEDT-TTF} molecules. Since the unit cell contains four molecules, four bands are found at the Fermi energy $E_F$ (depicted by the red dashed line).
Two are well separated from the Fermi energy whereas the upper band intersects $E_F$. From that, we conclude that \aeti\ is a metal with electrons as major carriers, which agrees with previous calculations\cite{Alemany12,Mori10} where also electron pockets were found and additionally small hole pockets. As discussed by Alemany {\it et al.}\cite{Alemany12}, the discrepancy can be ascribed to numerical uncertainties in the calculations since the difference are on the meV regime as well as to the (unrelaxed) used crystal structures leading to small deviations.

Fig.~\ref{fig:Bandstructure}(b) illustrates that a small indirect gap opens upon cooling in the charge-ordered state. Only 21~meV separate the bands from each other; direct gap amounts to 55~meV. This is in perfect agreement with the values from Ref.~\cite{Alemany12} with 30~meV and 60~meV, respectively.
From the dc measurements plotted in fig.~\ref{fig:Resistivity} we can extract a thermally activated behavior below $T\approx100$~K with an energy gap $\Delta_\rho=(60\pm 1)$~meV, in accordance with previous transport experiments [40~meV ($b$-axis) and 80~meV ($a$-axis)] and optical results of 75~meV \cite{Clauss10,Ivek11}. The total bandwidth $W$ is about 0.6~eV, which is in agreement with the values from the tight-binding model \cite{Mori10}.

The dispersion of the bands along the $c$-direction, i.e. perpendicular to the layers, is very weak and does not change considerable with temperature, except for a shift away from $E_F$.

\section{Optical conductivity}
\begin{figure*}
    \centering
    \includegraphics[width=0.7\textwidth]{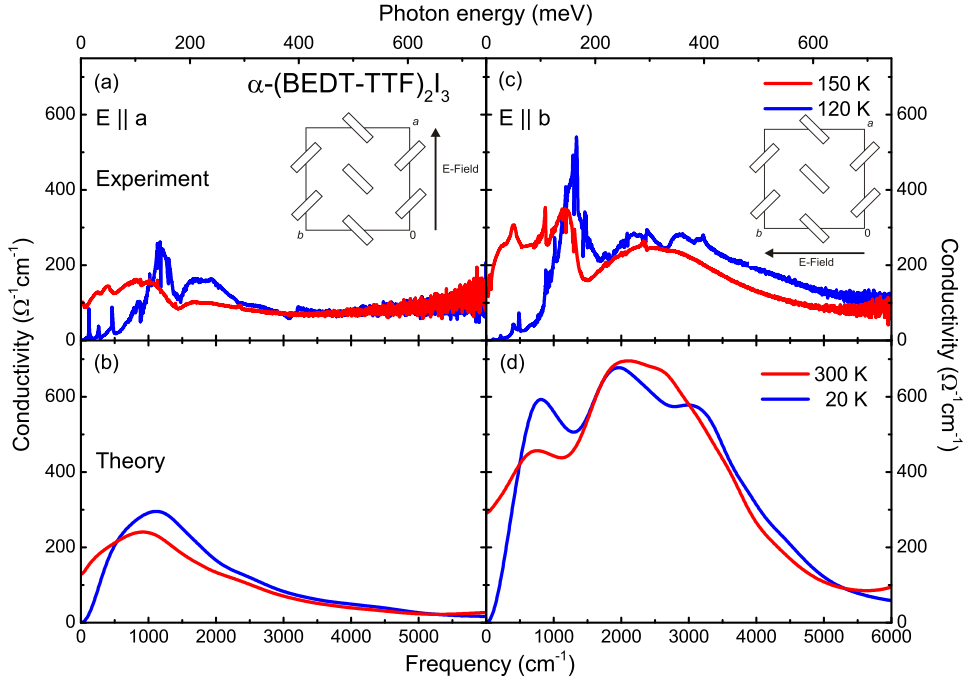}
    \caption{Measured optical conductivity $\sigma_1(\omega)$ of \aeti\ along the (a) $a$-direction and (c) $b$-direction for $T=150$~K (above $T_{\rm CO}$, red) and $T=120$~K (below $T_{\rm CO}$, blue). (b)~Theoretically calculated optical conductivity at ambient (red) and low temperatures (blue). (d) Corresponding simulated $\sigma_1(\omega)$ for the polarization  $E \parallel b$ at $T=300$~K (red) and 20~K (blue). The anisotropy, the temperature dependence as well as the spectral dependence of the conductivity is perfectly reproduced by the DFT calculations. Noteworthy, the calculations take into account solely electronic excitations. The inter- and intra-molecular vibrations present in the experimental spectra are unaccounted for by the calculations. \label{fig:Conductivity}}
\end{figure*}

Here, we focus on the theoretically and experimentally derived infrared conductivity.
The polarization dependent reflection data off the $ab$-plane of
a \aeti\ single crystal were collected between $T=300$~ and 20~K
according to the standard procedure\cite{Dressel04,Ivek11}.
Since the infrared optical properties of \aeti\ have been reported and discussed in many details previously\cite{Koch85,Sugano85,Meneghetti86,Yakushi87,Zelezny90,Dressel94,Clauss10,Yue10,Ivek10,Ivek11,Beyer16},
we can concentrate in this Letter on one typical spectrum in the metallic state ($T>T_{\rm CO}$) and one in the insulating state ($T<T_{\rm CO}$). The results are displayed in the upper panel of fig.~\ref{fig:Conductivity}.
The experiments are complemented by numerical simulations, where  we derived the optical response of \aeti\ from our DFT calculations after a self-consistent energy run with parameters and settings mentioned in the previous Section. The vertical electronic transitions were calculated on 200 equally distributed $k$-points in the Brillouin zone. The interband transitions were folded with Lorentzian functions of 100~meV width and the intraband transition with a 10~meV broad Gaussian function, respectively.

In fig.~\ref{fig:Conductivity} the conductivity $\sigma_1(\omega)$ measured in the frequency range from 100 to 6000~\cm\ for the polarization $E \parallel a$ and $E \parallel b$ at the temperatures $T=150$~K and 120~K is compared to the simulated optical conductivity for room temperature and  $T=20$~K.
The spectra taken along the $a$-direction [panel (a)]
exhibit an average conductivity of about $100~(\Omega{\rm cm})^{-1}$ above 2000~\cm. At lower frequencies it is dominated by a broad mid-infrared band which is disturbed by electron-molecular vibrational (emv)-coupled modes of the BEDT-TTF molecules.
In the limit $\omega \rightarrow 0$, the conductivity approaches $100~(\Omega{\rm cm})^{-1}$  for $T=150$~K in accord with the metallic character. Below $T_{\rm CO}$, $\sigma_1(\omega\rightarrow 0)$ drops to zero and an energy gap opens at about $\Delta_{\rm opt}=500$~\cm, corresponding to 60~meV, as a hallmark of the metal-insulator transition. It is remarkable how well this temperature evolution and spectral shape is rendered by the calculated spectra, plotted in fig.~\ref{fig:Conductivity}(b). The optical responses of the metallic and the insulating state are described very precisely. Also, the absolute values agree very well with the experimental findings, solely the vibrational modes are missing since they are not included in the model. The theoretical investigations allows us to assign the mid-infrared band to several interband transitions.

In contrast to the $a$-direction, the frequency-dependent conductivity along the $b$-axis
is  higher by at least a factor of 2, as seen in fig.~\ref{fig:Conductivity}(c). This agrees well with the fact that the orbital overlap is larger perpendicular to the stacks with respect to the stacking direction. The spectrum also reveals a broad mid-infrared band due to electronic origin with emv-coupled vibrational modes superimposed on top of it. In the metallic state the conductivity  below 1000~\cm\ is around $200~(\Omega{\rm cm})^{-1}$. Below the metal-insulator transition the low-frequency conductivity $\sigma_1(\omega)$ drops by several orders of magnitude, and already at $T=120$~K an energy gap has clearly been developed similar to the polarization $E\parallel a$. This is accompanied by a transfer of spectral weight from lower frequencies (i.e.\ $\omega/2\pi c = \nu < 1000$~\cm) to high frequencies. The calculated spectra in panel~(d) also shows a metallic behavior at room temperature. The mid-infrared band is composed of several excitations between electronic different bands. Due to the small broadening of 100~meV they can be distinguished, but should not be mixed up with the vibrational features of the  emv-coupled modes. At low temperatures the spectrum reveals an energy gap and the conductivity drops rapidly to zero. As aforementioned, the determined direct gap value is $\Delta_{\rm calc}\approx 55$~meV (440~\cm) which is obscured by the broadening of the interband transition.

\section{Discussion}
Our band structure calculations for \aeti\ at ambient conditions yield a metallic state with electrons as the majority charge carrier; the electron pockets are located around the S-point. Although the hole band reaches very close to $E_F$, we do not see any crossing. Previous calculations with relaxed molecules and unit-cell parameters have predicted a semi-metallic behavior with very small hole in addition to the large electron pockets\cite{Alemany12}. Since the bandwidth is rather small, any subtle change of the structure and temperature will shift $E_F$. Thus, from the theoretical side there remain some ambiguity, what the majority carriers are. Note, however, thermopower and Hall measurements unequivocally evidence hole conduction\cite{Bender84a,Pokhodnya87,Tajima00}.

From fig.~\ref{fig:Bandstructure} it becomes obvious that the contact point of the quasi Dirac cone fall
between the S- and X-point. Note that the evolution of the Dirac band structure is a delicate affair since small variations of the temperature as well as the strength of the applied pressure and its manner can influence the band dispersion\cite{Katayama06,Ishibashi06,Mori10,Alemany12,Suzumura12}. While in graphene the  energy range with linear dispersion extends over 1~eV\cite{CastroNeto09}, for \aeti\ it is less than 100~meV. Taking into account the smearing of the density of states at high temperature, the contribution of the Dirac
fermions to the optical conductivity is only possible at high pressures and very low temperatures. However, even in this case the Fermi energy intersects other parts of the band structure and, hence the Drude component of the ordinary charge carriers can obscure the view on the Dirac fermions. Very recently, extensive low-temperature investigations of the optical properties of \aeti\ under hydrostatic pressure have confirmed this conclusion\cite{Beyer16}.

Comparing our optical data with the theoretically determined conductivity spectra (fig.~\ref{fig:Conductivity}),
the excellent agreement is obvious in many respects:
The absolute value, the in-plane dc resistivity ration $\rho_a : \rho_b=2:1$, and also
the anisotropy of $\sigma_1(\omega)$ along the $a$- and $b$-direction is determined correctly.
But also the spectral shape for both polarizations and phases is reproduced very well.
Most important, our numerical simulation yield the transfer of spectral weight due to the phase transition at $T_{\rm CO}$ rather accurately.
The low-temperature insulating phase exhibits a band gap of $\Delta_{\rm calc}\approx 55$~meV, which agrees perfectly with the results from our optical and transport studies. Since electron-electron interactions
are only included to a certain fraction in DFT calculations, we have to question whether electronic correlations actually drive the transition into the charge-ordered state. At least they are not obvious from our optical spectra.

This appears to be different in the case of half-filled organic conductors, such as $\kappa$-(BEDT-TTF)$_2$\-Cu[N(CN)$_2$]\-Br$_x$Cl$_{1-x}$, where the optical spectra\cite{Faltermeier07,Merino08,Dumm09,Dressel09,Dressel10} could be well reproduced by
a combination of density functional theory calculations and the dynamical mean field theory approach\cite{Ferber14}. Tuning the bandwidth $W$ by pressure, the dimerized Mott insulator is transformed to a Fermi-liquid and superconductor\cite{Yasin11,Dressel11}. The evolution of the experimentally observed spectra by chemical pressure via the Br-content $x$ is well reproduced by variation of the effective correlations $U/W$. The double-natured origin of the mid-infrared peak in $\sigma_1(\omega)$ suggested from the experimental data\cite{Faltermeier07}
was confirmed: an interband transition which is unaffected by electronic correlations and a correlation-induced intraband contribution. The latter one scales with $U$ but is to the charge transfer within an BEDT-TTF dimer present in the $\kappa$-phase.  It would be interesting to extend these theoretical investigations to non-dimerized sytems, which are commonly described by the extended Hubbard model with on-site and inter-site Coulomb repulsion, $U$ and  $V$, being of relevance\cite{Kino95,Kino96,Seo04}.

\section{Conclusions}
Our comparative theoretical and experimental study on the optical properties of \aeti\ at the metal-insulator transition reveals a great deal of agreement.  From our DFT calculations of the electronic band structure, we
derive frequency dependent conductivity in the metallic and charge-ordered state that match the optical spectra measured the reflectivity in all crystallographic directions. We can precisely reproduce the temperature-dependent evolution and shape, as well as the shift in spectral weight and opening of the optical gap. We do not observe any indications of correlations in our optical spectra of \aeti.
This endorses the idea that correlation effects are not as important to drive the system into the charge-ordered phase as previously presumed.

\begin{acknowledgments}
We would like to thank E. Rose, D. Wu and S. Zapf for useful discussions as well
as G. Untereiner for technical support. We thank D. Schweitzer for providing the single crystals. Funding by the Deutsche Forschungsgemeinschaft (DFG) and the Carl-Zeiss Stiftung is acknowledged.
\end{acknowledgments}


\end{document}